# A Primary Exploration to Quasi-Two-Dimensional Rare-Earth Ferromagnetic Particles: Holmium-Doped MoS$_2$ Sheet as Room-Temperature Magnetic Semiconductor

Xi Chen and Zheng-Zhe Lin*

*Department of Applied Physics, School of Physics and Optoelectronic Engineering, Xidian University, Xi'an 710071, China*

*Corresponding Author. E-mail address:   linzhengzhe@hotmail.com

**Abstract** – Recently, two-dimensional materials and nanoparticles with robust ferromagnetism are even of great interest to explore basic physics in nanoscale spintronics. More importantly, room-temperature magnetic semiconducting materials with high Curie temperature is essential for developing next-generation spintronic and quantum computing devices. Here, we develop a theoretical model on the basis of density functional theory calculations and the Ruderman-Kittel-Kasuya-Yoshida theory to predict the thermal stability of two-dimensional magnetic materials. Compared with other rare-earth (dysprosium (Dy) and erbium (Er)) and 3$d$ (copper (Cu)) impurities, holmium-doped (Ho-doped) single-layer 1H-MoS$_2$ is proposed as promising semiconductor with robust magnetism. The calculations at the level of hybrid HSE06 functional predict a Curie temperature much higher than room temperature. Ho-doped MoS$_2$ sheet possesses fully spin-polarized valence and conduction bands, which is a prerequisite for flexible spintronic applications.

## 1. Introduction

In the study of spintronics, people keep seeking effective techniques to exploit the electron spin for logic devices and data storage because it is much easier to change the state of electron spin than to change charge. However, at present the generation and manipulation of fully spin-polarized currents are still key challenges. In future spintronics, magnetic semiconductors would play an important role because both charge and spin could be effectively manipulated by them. By controlling electron spin degree of freedom in magnetic semiconductors, the logic operations of semiconductor devices and the storage functions of magnetic materials could be integrated on a single chip. Using magnetic semiconductor devices, greatly enhanced computing speed could be achieved as well as reduced power consumption (Wolf, Awschalom et al. 2001). In the past decades, scientists have been trying to combine semiconductors with magnetic materials. The major technology to prepare magnetic semiconductors is to dope magnetic elements into traditional semiconductors (Dietl 2010, MacDonald, Schiffer et al. 2005, Matsumoto, Murakami et al. 2001, Ohno, Chiba et al. 2000). However, the Curie temperature of existing magnetic semiconductors (e.g. EuS (von Molnar and Methfessel 1968), (In, Mn)As (Munekata, Ohno et al. 1989) and (Ga, Mn)As (Chen, Yang et al. 2011, Ohno, Shen et al. 1996)) is still far below room temperature. Recently, the Curie temperature of typical magnetic semiconductors such as (Ga, Mn)As has been improved to about 200 K (Maccherozzi, Sperl et al. 2008, Nie, Chin et al. 2013, Song, Sperl et al. 2011). Until now, room-temperature magnetic semiconductor is still a major issue for the realization of semiconductor spintronic devices.

In recent years, two-dimensional materials with intrinsic and robust ferromagnetism are of great interest. Magnetism in two-dimensional materials may have valuable applications in quantum computation and logic operation. However, the ferromagnetic ordering in single atomic layer has been particularly elusive. According to the Mermin-Wagner theorem (Mermin and Wagner 1966), the long-range magnetic order would be strongly suppressed in two-dimensional materials. However, just recently, two-dimensional $Cr_2Ge_2Te_6$ (Gong, Li et al. 2017) and $CrI_3$ (Huang, Clark et al. 2017) van der Waals crystals are found to possess intrinsic ferromagnetism.

Room-temperature ferromagnetism of two-dimensional $Fe_2Si$ (Sun, Zhuo et al. 2017) and nitride MXenes (Kumar, Frey et al. 2017) has been theoretically proposed. These new results imply that magnetic ground state might be retained in a single atomic layer. The development of two-dimensional magnetic materials, which breaks the limit of Mermin-Wagner theorem, is of even greater significance than general three-dimensional magnetic materials. The discovery of two-dimensional magnetic materials opens the door for new opportunities to explore ultra-compact spintronics.

Transition metal dichalcogenides, composed of atomic monolayers with weak interlayer bonding, are a class of materials with various structures and properties. The breakthroughs in the preparation and manipulation of two-dimensional transition metal dichalcogenides (Coleman, Lotya et al. 2011, Cunningham, Lotya et al. 2012, Lee, Zhang et al. 2012, Mak, Lee et al. 2010, Radisavljevic, Radenovic et al. 2011, Zhan, Liu et al. 2012) have enabled new possibilities for spintronic applications as well as fundamental studies. Some transition metal dichalcogenides are $d^0$ (e.g. $TiS_2$) or band insulators (e.g. $MoS_2$) in which an even number of $d$ electrons completely fills the valence band. Transition metal impurities with extra $3d$ electrons have been proposed to inducing magnetism in single-layer $MoS_2$ (Cheng, Zhu et al. 2013, Yue, Chang et al. 2013). A $3d$ transition metal impurity with electron configuration $3d^N4s^2$ has $N$-4 additional valence electrons as compared to Mo ($4d^54s^1$). For a $3d$ transition metal impurity located at Mo vacancy, six electrons occupy the valence band of $MoS_2$ and the remaining $N$-4 electrons occupy the $3d$ orbitals of impurity atom in high-spin state (Cheng, Zhu et al. 2013, Yue, Chang et al. 2013). According to previous calculations, transition metal impurities, such as Mn, Fe, Co, Ni and Cu, possess magnetic moments of 1, 2, 3, 4 and 5 $\mu_B$ in $MoS_2$ sheet, respectively (Cheng, Zhu et al. 2013, Xia, Guo et al. 2016, Yue, Chang et al. 2013). Recent experiments have identified the magnetism of Mn-doped $MoS_2$ (Wang, Sun et al. 2016) and Cu-doped $MoS_2$ (Xia, Guo et al. 2016). Compared with $3d$ transition metal ($d^5$ configuration as the highest spin state), rare-earth elements ($f^7$ configuration as the highest spin state), with more localized $4f$ orbitals, would induce much stronger magnetism in $MoS_2$. To further enhance the Curie temperature, we then consider theoretically studying rare-earth impurities in MoS$_2$, which could provide beneficial guidance for realizing room-temperature magnetic conductor. Furthermore, realizing stable magnetism in two-dimensional systems is of more profound significance than in ordinary three-dimensional systems. Thus we consider studying the ferromagnetic ordering of rare-earth magnetic impurities in single-layer MoS$_2$, which would be beneficial for furure research on two-dimensional magnetism.

In this work, the magnetism and electronic properties of rare-earth-doped single-layer MoS$_2$ are studied using density functional theory (DFT) calculations and the Ruderman-Kittel-Kasuya-Yoshida (RKKY) theory (Kittel 1987, Ruderman and Kittel 1954, Yosida 1957). Holmium (Ho) substitutional impurities in 1H-MoS$_2$ sheet are found to possess strong magnetism with Curie temperature much higher than room temperature. By means of mean-field theory and Monte Carlo simulations based on Ising model, the reliability of our results is sufficiently demonstrated. The Curie temperature of Ho-doped MoS$_2$ mainly depends on Ho ratio, but less depends on the positions of Ho impurity atoms. By contrast, other rare-earth (dysprosium (Dy) and erbium (Er)) and 3d (Cu) impurities present much lower Curie temperature. With both fully polarized electrons and holes, Ho-doped MoS$_2$ possesses a very promising candidate for flexible spintronic applications.

## 2. Computational details

First-principles calculations are performed within spin-polarized DFT by using Vienna *ab initio* simulation package (VASP) (Kresse and Hafner 1993). Projector-augmented wave (PAW) method (Blöchl 1994) is used to account electron-ion interactions. A plane-wave basis set with kinetic energy cutoff of 400 eV is used to expand the wave functions. To validate the calculations, higher plane-wave energy cutoff (up to 650 eV) is also tried to reproduce similar results. To deal with the strong correlation of d- and f-orbital electrons, the exchange-correlation potential is represented by the hybrid Heyd-Scuseria-Ernzerhof (HSE06) functional (Heyd,

Scuseria et al. 2003, Heyd, Scuseria et al. 2006). To save computation time, the generalized gradient approximation (GGA) of Perdew-Burke-Ernzerhof (PBE) (Perdew, Burke et al. 1996) is used as preprocessing, providing initial wave functions for HSE06 calculations. In the PBE precalculations, the Hubbard $U$ correction is employed in Dudarev's rotationally invariant formalism (Dudarev, Botton et al. 1998), with the $U$ values for Mo and rare-earth elements taken from Ref. (Topsakal and Wentzcovitch 2014, Williamson, Li et al. 2017).

In all the calculations, 4×4 supercell of 1H-MoS$_2$ single layer is used as simulation system. The Brillouin zone is sampled by using 2×2×1 Γ-centered Monkhorst-Pack grid. A Gaussian smearing with a width of $\sigma = 0.05$ eV is used. The convergence of the total energy is considered to be achieved until two iterated steps with energy difference less than $10^{-5}$ eV. The replicas of MoS$_2$ layers are separated by a large spacing of 16 Å to avoid interlayer interactions. Geometry relaxations are performed until the Hellmann-Feynman forces acting on each atom are less than 0.01 eV/Å. For the calculations of ferromagnetic (FM) and antiferromagnetic (AFM) states, different initial magnetic moments are tested to avoid being trapped in a local minimum.

## 3. Results and discussion

### 3.1 *Basic properties*

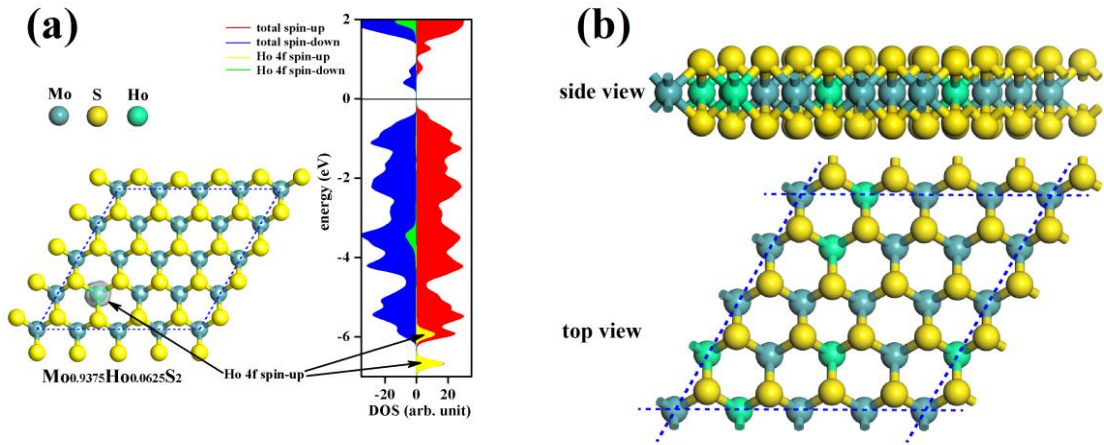

Fig. 1 (a) The structure and DOS of 4×4 MoS$_2$ simulation system with a single Ho atom (with empirical formula Mo$_{0.9375}$Ho$_{0.0625}$S$_2$). The charge density Ho 4f orbitals is shown by the isosurfaces of 0.05 e/Å$^3$. (b) The top and side views of simulation model for Ho-substituted MoS$_2$ sheet.

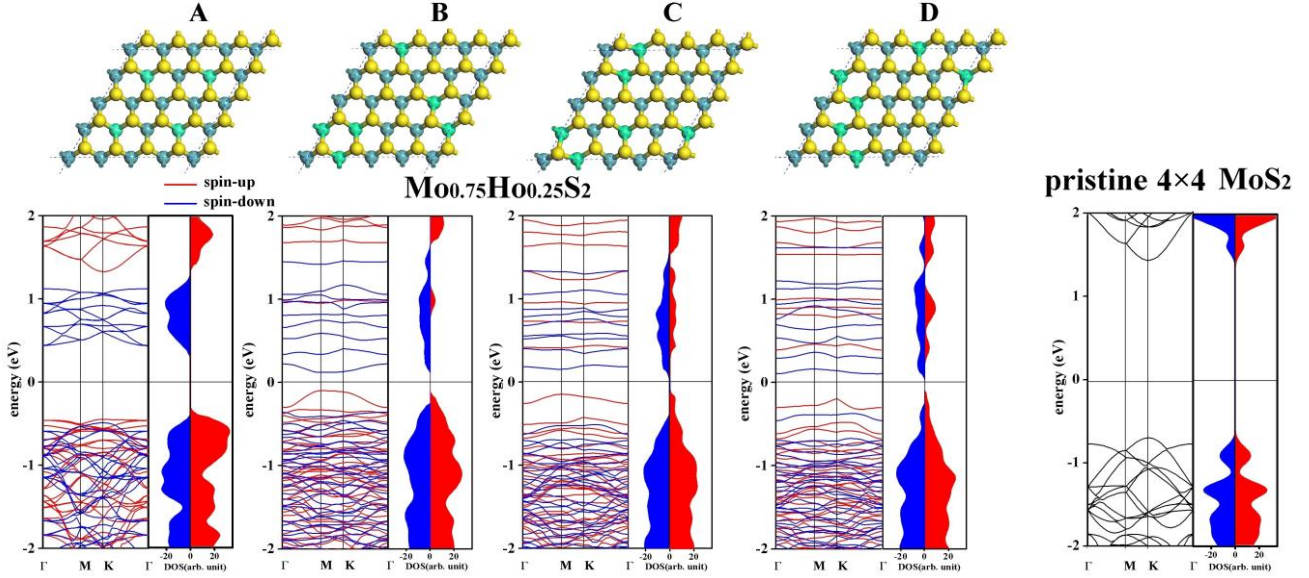

Fig. 2 The energy bands and DOS of four random structures (**A**, **B**, **C** and **D**) of Ho-substituted MoS$_2$ sheet (with empirical formula Mo$_{0.75}$Ho$_{0.25}$S$_2$). In the above figures, the simulation cells are shown in dashed lines. The energy bands and DOS of pristine 4×4 MoS$_2$ are shown on the right. The Fermi level is set to zero.

The magnetic moments of rare-earth impurities in MoS$_2$ originate from extra electrons filled into $f$ orbitals. When a rare-earth atom with $x$ valence electrons substitutes a Mo ($4d^5 5s^1$, $x=6$) atom, it contributes $x-6$ extra electrons to $f$ orbitals. In lanthanide elements, Ho ($4f^{11} 6s^2$, $x=13$) atom could form $f^7$ configuration with maximum atomic spin $S=7/2$. Therefore, we mainly focus on Ho substitutional impurities in MoS$_2$ sheet. Fig. 1(a) shows the density of states (DOS) of the 4×4 MoS$_2$ simulation system with a single Ho atom (with empirical formula Mo$_{0.9375}$Ho$_{0.0625}$S$_2$). Obviously, the occupied Ho 4f orbitals are highly spin-polarized and deeply buried in 6~7 eV under the Fermi level. The calculated atomic spin of Ho impurity atom is just $S=7/2$. Then, to systematically investigate the magnetic properties of Ho impurity atoms in single-layer MoS$_2$, a simulation model is set up by randomly introducing four Ho substitutional atoms into a 4×4 supercell of 1H-MoS$_2$ (Fig. 1(b)). We investigate four different structures (**A**, **B**, **C** and **D**), with their energy bands and DOS plotted in Fig. 2 (with empirical formula Mo$_{0.75}$Ho$_{0.25}$S$_2$). Here, in the system **A** the Ho impurity atoms are arranged periodically. In the system **B**, **C** and **D**, Ho impurity atoms are arranged randomly. These systems are all magnetic semiconductors with FM ground states, in which every Ho impurity atom possesses parallel spin $S=7/2$. It can be seen that these energy bands all have high spin polarization, with the spin-up bands highly shifted relative to the spin-down bands. The valence and conduction bands near the Fermi level possess opposite spin orientations. So, the holes near the valence band minimum (VBM) are spin-up, and the electrons near the conduction band minimum (CBM) are spin-down. We also exhibit the band profile of 4×4 pristine MoS$_2$ lattice (on the right of Fig. 2) for comparison. For the system **A**, **B C** and **D**, the spin-up band gaps are in the range of 0.56~1.79 eV, and the spin-down band gaps are in the range of 0.48~0.93 eV, which are all smaller than the band gap of pristine MoS$_2$ (2.13 eV via HSE06 calculation, 1.8~1.9 eV via experiments (Mak, Lee et al. 2010, Splendiani, Sun et al. 2010)). For their valence bands, the DOS profiles keep some similar feature to pristine MoS$_2$. But the conduction bands are highly spin-polarized. In the band profiles, it can be seen that the substitution of Mo atoms by Ho results in a transition from direct to indirect band gap.

3.2 *Carrier polarization*

To investigate the transport properties of carriers, carrier concentration and spin polarizability are evaluated. In the conduction bands, the spin-up/spin-down electron concentration reads

$$n_{h\uparrow\downarrow} = \frac{1}{A}\int_{E>\text{CBM}\uparrow\downarrow} f(E,E_F) D_{\uparrow\downarrow}(E) dE \quad (1)$$

with the integral over energy $E$ above the spin-up/spin-down CBM. Here, $D_{\uparrow\downarrow}(E)$ denotes the spin-up/spin-down DOS at energy $E$, $E_F$ denotes the Fermi energy, $T$ denotes the temperature, and $A$ denotes the area of simulation cell. $f(E,E_F) = [\exp((E-E_F)/kT)+1]^{-1}$ denotes the Fermi-Dirac distribution. In the valence bands, the spin-up/spin-down hole concentration reads

$$n_{e\uparrow\downarrow} = \frac{1}{A}\int_{E<\text{VBM}\uparrow\downarrow}(1-f(E,E_F))D_{\uparrow\downarrow}(E)dE \qquad (2)$$

with the integral over energy $E$ below the spin-up/spin-down VBM. The spin polarizability of holes and electrons reads $P_h=|n_{h\uparrow}-n_{h\downarrow}|/(n_{h\uparrow}+n_{h\downarrow})$ and $P_e=|n_{e\uparrow}-n_{e\downarrow}|/(n_{e\uparrow}+n_{e\downarrow})$, respectively.

The calculations of spin-up/spin-down carrier concentration show that the electrons and holes in these four systems are almost fully polarized. For the system **A**, the polarizability of electrons/holes is 100% / 93%, respectively. For the system **B**, **C** and **D**, the polarizability of electrons and holes are all nearly 100%. These results show that Ho-doped MoS$_2$ system possesses spin-up holes as well as spin-down electrons. Such dual character may be utilized for special spintronic devices. Table 1 lists the carrier concentration in the system **A**, **B**, **C** and **D**. It is worth noting that Ho impurities in MoS$_2$ greatly decrease the band gap. For electrons in conduction bands and holes in valence bands, this results in much larger Fermi-Dirac distribution than pristine MoS$_2$. With much larger carrier concentration than pristine MoS$_2$ ($1.5\times10^{-4}$ cm$^{-2}$ in our HSE06 calculation), these systems should have nice conductivity. Since there are no impurity levels in the band gap, the concentration of holes should be equal to the concentration of electrons.

| system | electrons | | holes | |
|---|---|---|---|---|
| | concentration $n_e$ (cm$^{-2}$) | spin polarizability $P_e$ | concentration $n_h$ (cm$^{-2}$) | spin polarizability $P_h$ |
| **A** | $5.5\times10^8$ | 100 | $5.5\times10^8$ | 93 |
| **B** | $4.8\times10^{12}$ | 100 | $4.8\times10^{12}$ | 99 |
| **C** | $2.3\times10^{12}$ | 100 | $2.3\times10^{12}$ | 100 |
| **D** | $4.9\times10^{12}$ | 100 | $4.9\times10^{12}$ | 100 |

Table 1 The carrier concentration and spin polarizability in the system **A**, **B**, **C** and **D**.

### 3.3 *RKKY theory and the thermal stability of magnetism*

To investigate the magnetic ordering of rare-earth impurity atoms in MoS$_2$ sheet, we employ the RKKY Hamiltonian (Kittel 1987, Ruderman and Kittel 1954, Yosida 1957) for local magnetic moments to describe the indirect exchange coupling through conduction electrons. According to Kasuya's theory (Kasuya 1956), the interaction between the conduction electrons and an impurity atom located at **R** reads

$$\mathcal{H} = -\frac{J}{N} \sum_{\mathbf{k}\mathbf{k}'\mu\mu'} e^{i(\mathbf{k}'-\mathbf{k})\cdot\mathbf{R}} \mathbf{S} \cdot \boldsymbol{\sigma}_{\mu\mu'} C^+_{\mathbf{k}\mu} C^+_{\mathbf{k}'\mu'}. \quad (3)$$

Here, $C_{\mathbf{k}\mu}$ is the annihilation operator of Blöch state with wave vector **k** and spin $\mu$. **S** is the spin of impurity atom. $\boldsymbol{\sigma}$ is the Pauli matrix. $J$ is the exchange integral. $N$ is the number of unit cells. For two magnetic rare-earth atoms with spin $\mathbf{S}_i$ ($\mathbf{S}_j$) located at $\mathbf{R}_i$ ($\mathbf{R}_j$), their second-order interaction (Kittel 1987, Li 2002) reads

$$H_{RKKY}(\mathbf{R}_i - \mathbf{R}_j) = -\sum_{\mathbf{k}\mathbf{k}'\mu\mu'} \frac{\langle \mathbf{k}\mu | \mathcal{H} | \mathbf{k}'\mu' \rangle \langle \mathbf{k}'\mu' | \mathcal{H} | \mathbf{k}\mu \rangle}{\varepsilon(\mathbf{k}') - \varepsilon(\mathbf{k})}$$

$$= -\mathbf{S}_i \cdot \mathbf{S}_j \frac{J^2}{N^2} \sum_{\substack{\varepsilon(\mathbf{k}) < E_F \\ \varepsilon(\mathbf{k}') > E_F}} \frac{e^{i(\mathbf{k}'-\mathbf{k})\cdot(\mathbf{R}_i-\mathbf{R}_j)}}{\varepsilon(\mathbf{k}') - \varepsilon(\mathbf{k})}, \quad (4)$$

where $\varepsilon(\mathbf{k})$ is the energy of Blöch state $|\mathbf{k}\mu\rangle$. The summation is taken over $\varepsilon(\mathbf{k})$ below the Fermi level $E_F$ and $\varepsilon(\mathbf{k}')$ above $E_F$. For two-dimensional lattice with direct band gap, Eq. (4) reads (for detail derivation see Appendix)

$$H_{RKKY}(\mathbf{R}_i - \mathbf{R}_j) = -\mathbf{S}_i \cdot \mathbf{S}_j \frac{J^2 A_0^2}{(2\pi)^4} \int_{\substack{\varepsilon(\mathbf{k}) < E_F \\ \varepsilon(\mathbf{k}') > E_F}} \frac{e^{i(\mathbf{k}'-\mathbf{k})\cdot(\mathbf{R}_i-\mathbf{R}_j)}}{\varepsilon(\mathbf{k}') - \varepsilon(\mathbf{k})} dk_x dk_y dk'_x dk'_y \quad (5)$$

$$= -\mathbf{S}_i \cdot \mathbf{S}_j \frac{J^2 A_0^2}{(2\pi)^2} \int_{\substack{\varepsilon(\mathbf{k}) < E_F \\ \varepsilon(\mathbf{k}') > E_F}} \frac{\mathrm{k}\mathrm{k}' J_0(\mathrm{k}|\mathbf{R}_i - \mathbf{R}_j|) J_0(\mathrm{k}'|\mathbf{R}_i - \mathbf{R}_j|)}{\varepsilon(\mathrm{k}') - \varepsilon(\mathrm{k})} d\mathrm{k}d\mathrm{k}'. \quad (6)$$

Here, $A_0$ is the area of unit cell. $J_0$ is the zero-order Bessel function. For single-layer MoS$_2$, the integral in Eq. (6) is performed numerically on the basis of energy band calculation, obtaining the function $H_{RKKY}(\mathbf{R}_i - \mathbf{R}_j)$ with changing interatomic distance $\mathbf{R}_i - \mathbf{R}_j$. With the Hamiltonian defined by Eq. (5), we employ Ising model to simulate a MoS$_2$ sheet with random substitution of rare-earth atoms. The rare-earth atomic spins $\mathbf{S}_i$ are sampled in thermal equilibrium by the Metropolis Monte Carlo method. The Curie temperature is evaluated by the magnetic moment and specific heat changing with temperature.

The thermal stability of ferromagnetism and Curie temperature of FM Ho-doped MoS$_2$ is a key physical property for its spintronic applications. To investigate the change of magnetic moment against temperature, we employ the RKKY Hamiltonian in Eq. (5). For a system with magnetic moments $\mathbf{S}_1$, $\mathbf{S}_2$, … , $\mathbf{S}_N$ located at $\mathbf{R}_1$, $\mathbf{R}_2$, … , $\mathbf{R}_N$, respectively, the RKKY Hamiltonian reads $H_{RKKY} = -\sum_{i<j} \mathbf{S}_i \cdot \mathbf{S}_j \frac{J^2 A_0^2}{(2\pi)^4} I(\mathbf{R}_i - \mathbf{R}_j)$. Before performing the Monte Carlo simulations, the integral $I(\mathbf{R}_i - \mathbf{R}_j)$ is numerically determined using Eq. (6) and the energy bands of pristine MoS$_2$ at the level of HSE06. The numerical result of $I(\mathbf{R}_i - \mathbf{R}_j)$ is plotted in Fig. 3(a). Then, the coupling $J$ in Eq. (5) is obtained using the energy difference between FM and AFM states of Ho-doped MoS$_2$. Fig. 3(b) shows the FM and AFM states of the 4×4 supercell of MoS$_2$ with four Ho impurities arranged periodically. For one Ho atom, we can see six nearest Ho atoms. In comparison with the FM state, in the AFM state four of the six nearest neighbors change the magnetic moments. So, from the FM to AFM state, the energy change per Ho atom is $\Delta E = E_{AFM} - E_{FM} = 4S^2 \frac{J^2 A_0^2}{(2\pi)^4} I(\mathbf{R}_i - \mathbf{R}_j)$. Here $S=7/2$ is the Ho spin, $\mathbf{R}_i - \mathbf{R}_j = 6.36$ Å is the interatomic distance, and $A_0 = 8.78$ Å$^2$ is the area of MoS$_2$ primitive cell. According to DFT calculation which gives $\Delta E = E_{AFM} - E_{FM} = 0.307$ eV (per Ho atom), we then have $J = 14.30$ eV. The above analysis provides quantitative parameters for the following Monte Carlo simulations.

Then, to investigate the relation between Ho impurity ratio and the Curie temperature, Monte Carlo simulations are performed using Ising model with the RKKY Hamiltonian. The simulation system includes $n$ Ho impurities and 256-$n$ Mo atoms, with $n$=16, 32, 48 and 64 corresponding to 6.25, 12.5, 18.75 and 25 % Ho ratio, respectively. For each Ho ratio, the Ho atoms are arranged randomly in a 16×16 MoS$_2$ lattice. The simulation temperature $T$ ranges from 100 K to 3000 K. In the simulations, the magnetic moment $<M>$ and susceptibility $\chi=(<M^2>-<M>^2)/kT$ are calculated. Fig.

3(c) represents an example of <*M*> and $\chi$ changing with temperature. At the Curie temperature $T_c$, the magnetic susceptibility $\chi$ reaches the maximum and the magnetic moment <*M*> sharply decreases. For each Ho ratio, the simulations are performed 40 times at each temperature with randomly arranged Ho atoms. The results (shown by squares in Fig. 3(d)) represent a rising Curie temperature $T_c$ with increasing Ho ratio. For 6.25, 12.5, 18.75 and 25 % Ho, the average Curie temperatures are about $T_c$=280, 750, 1820 and 1960 K, respectively. For Ho ratio higher than 6.25 %, the Curie temperature $T_c$ is even much higher than room temperature, indicating the possibility of Ho-doped MoS$_2$ as room-temperature magnetic semiconductor. It is worth noting that in the simulations at a specific Ho ratio with random Ho position, the variance of Curie temperature is in a range of ±100 K. This result indicates that the Curie temperature of Ho-doped MoS$_2$ mainly depends on Ho ratio, but less depends on Ho impurity position.

For comparison, we also calculate the Curie temperature using mean-field theory. For a system with Ho impurities arranged periodically, e.g. Fig. 3(b), the Curie temperature predicted by mean-field theory is $T_c = \frac{2zS(S+1)}{3k} \frac{J^2 A_0^2}{2(2\pi)^4} I(\mathbf{R}_i - \mathbf{R}_j)$, where $z$=6 is the number of Ho nearest neighbor. For the systems with each Ho impurity in 4×4, 3×3 and 2×2 MoS$_2$ supercell (corresponding to 6.25, 11.1 and 25 % Ho ratio), the calculated $T_c$=16, 114 and 2290 K (shown by crosses in Fig. 3(d)), respectively. It can be seen that for low Ho ratio (6.25 and 11.1 %) the $T_c$ calculated using mean-field theory is much lower than Monte Carlo simulations (Fig. 3(d)). This result can be explained that, in the model of mean-field theory the Ho impurities are arranged periodically. By contrast, in Monte Carlo simulations the Ho impurities in random positions may partially aggregate and enhance the magnetic coupling. So, the periodical model just provides the lower limit of $T_c$. But the Monte Carlo simulations provide more actual $T_c$.

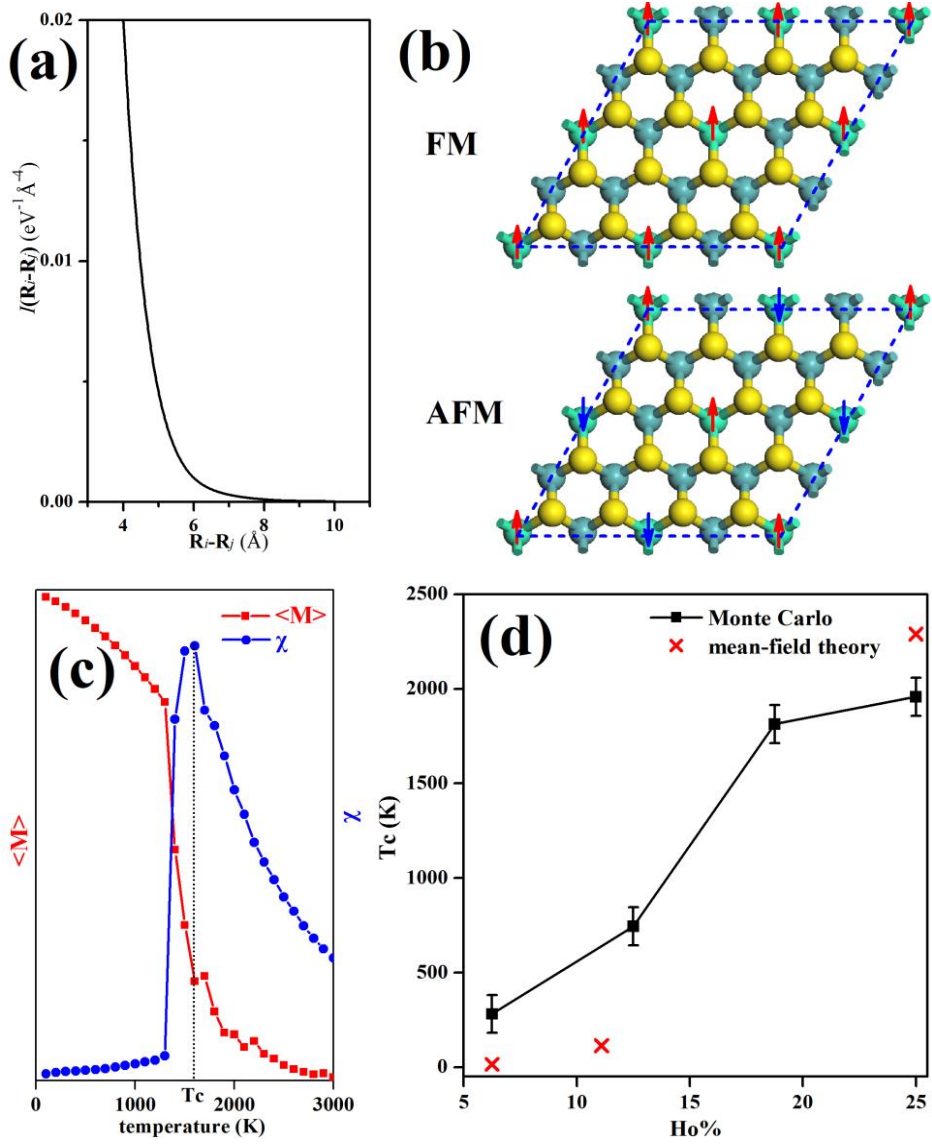

Fig. 3 (a) The RKKY-related integral $I(\mathbf{R}_i-\mathbf{R}_j)$ (Eq. (6)) versus the interatomic distance $\mathbf{R}_i$-$\mathbf{R}_j$. (b) The FM and AFM states of the 4×4 supercell of $MoS_2$ with four Ho impurities arranged periodically. The local magnetic moments of Ho impurities are shown by up and down arrows. The simulation cells are shown in dashed lines. (c) An example of Monte Carlo simulation. The magnetic moment $<M>$ and susceptibility $\chi$ are plotted against temperature. $T_c$ is the Curie temperature. (d) The Curie temperature $T_c$ changing with Ho impurity ratio. The black squares denote the average values from Monte Carlo simulations. The red crosses denote the values from mean-field theory.

3.4 *Other rare-earth or transition-metal impurities*

Finally, we extend the study to investigate the effect of other rare-earth impurities in $MoS_2$. We focus on Dy ($4f^{10}6s^2$, $x=12$) and Er ($4f^{12}6s^2$, $x=14$) impurities which form $f^6$ and $f^8$ configurations in $MoS_2$, respectively. They are both with atomic spin of $S=3$ in $MoS_2$. The Curie temperatures are roughly estimated using mean-field theory. We perform DFT calculations for the system shown in Fig. 3(b), obtaining the energy difference between the FM and AFM state ($\Delta E = E_{AFM} - E_{FM} = 0.039$ (0.074) eV per Dy (Er) atom, respectively). According to above theory, the coupling $J = 5.95$ (8.19) eV for Dy (Er) impurity, respectively. For the system with 25 % Dy (Er) impurity ratio, by the mean-field theory the Curie temperature is calculated to be $T_c$=300 (570) K, respectively, which is much lower than that of 25 % Ho (2290 K). We also consider Cu ($3d^{10}4s^1$, $x=11$) impurity (which forms $d^5$ configuration in $MoS_2$) with atomic spin $S=5/2$. The calculation gives $\Delta E = E_{AFM} - E_{FM} = 0.026$ eV, $J = 5.83$ eV and $T_c$=210 K of 25 % Cu doping. It can be judged that for Dr, Er or Cu ratio lower than 25 %, the Curie temperature would be even lower than room temperature. Overall, Dr, Er and Cu impurities in $MoS_2$ are less suitable than Ho to realize room-temperature magnetic semiconductor.

## 4. Conclusions

In this work, single-layer $MoS_2$ sheet with Ho impurities is proposed as promising room-temperature magnetic semiconductor. By means of mean-field theory and Monte Carlo simulations based on Ising model, the existence of Ho-doped $MoS_2$ as room-temperature magnetic semiconductor is sufficiently demonstrated. Ho impurity atoms in $MoS_2$ present maximum spin $S=7/2$ ($f^7$ configuration). Combining DFT calculations and the RKKY theory, the magnetic coupling parameters are obtained. Then, the Curie temperature and temperature-dependent magnetic susceptibility are derived using Monte Carlo simulations based on Ising model and the RKKY Hamiltonian. According to the results, $MoS_2$ sheet with more than 12.5 % Ho impurity is suggested as room-temperature magnetic semiconductor, with Curie temperature higher than 750 K. At a Ho impurity ratio of 25 %, both Monte Carlo simulation and mean-field theory predict a Curie temperature higher than 1900 K. The Curie temperature of Ho-doped $MoS_2$ mainly depends on Ho ratio, but less depends on the positions of Ho impurity atoms.

For various kinds of randomly Ho-doped MoS$_2$ sheets, the valence and conduction bands all possess opposite spin orientations. The electrons in conduction bands and the holes in valence bands are almost fully spin-polarized, with carrier concentration of ~10$^3$ times larger than pristine MoS$_2$. With both fully spin-polarized electrons and holes, Ho-doped MoS$_2$ possesses a prerequisite property for flexible spintronic applications.

In contrast with Ho, Dy, Er and Cu impurities ($f^6$, $f^8$ and $d^5$ configurations, with atomic spin $S$=3, 3 and 5/2, respectively) present much lower Curie temperature. With a magnetic coupling of 42 / 57 / 41 % of Ho, Dy / Er / Cu impurities have a Curie temperature of about 13 / 25 / 9 % of Ho, respectively. Overall, Ho is much better than other rare-earth or 3$d$ transition metal for realizing room-temperature magnetism in MoS$_2$.

**Compliance with Ethical Standards:**

**Funding**: This study was funded by the National Natural Science Foundation of China (Grant No. 11304239) and the 111 Project (B17035).

**Conflict of Interest**: The authors declare that they have no conflict of interest.

**Appendix**

This appendix explains the analytical derivation of Eq. (5) and (6). The summation of Eq. (4) takes over the $k$-points in Brillouin zone. For a two-dimensional system with $N$ unit cells, the phase volume of every $k$-point is $(2\pi)^2/NA_0$, where $A_0$ is the area of one unit cell. The summation $\sum_{\mathbf{k}}...$ over the Brillouin zone can be replaced as $\frac{NA_0}{(2\pi)^2}\int...dk_x dk_y$. Thus Eq. (4) can be written as

$$H_{RKK}(\mathbf{R}_i - \mathbf{R}_j) = -\mathbf{S}_i \cdot \mathbf{S}_j \frac{J^2}{N^2} \sum_{\substack{\varepsilon(\mathbf{k}) < E_F \\ \varepsilon(\mathbf{k'}) > E_F}} \frac{e^{i(\mathbf{k'}-\mathbf{k})\cdot(\mathbf{R}_i-\mathbf{R}_j)}}{\varepsilon(\mathbf{k'}) - \varepsilon(\mathbf{k})}$$

$$= -\mathbf{S}_i \cdot \mathbf{S}_j \frac{J^2 A_0^2}{(2\pi)^4} \int_{\substack{\varepsilon(\mathbf{k}) < E_F \\ \varepsilon(\mathbf{k'}) > E_F}} \frac{e^{i(\mathbf{k'}-\mathbf{k})\cdot(\mathbf{R}_i-\mathbf{R}_j)}}{\varepsilon(\mathbf{k'}) - \varepsilon(\mathbf{k})} dk_x dk_y dk'_x dk'_y .$$

Then, $\varepsilon(\mathbf{k})$ near VBM or CBM is approximated by the effective mass, which reads $\varepsilon(\mathbf{k}) \approx E_{VBM} - m^*_{VBM} (\mathbf{k} - \mathbf{k}_{gap})^2$ for $\varepsilon(\mathbf{k}) < E_F$ and $\varepsilon(\mathbf{k}) \approx E_{CBM} + m^*_{CBM} (\mathbf{k} - \mathbf{k}_{gap})^2$ for $\varepsilon(\mathbf{k}) > E_F$. Here, $E_{VBM}$ ($E_{CBM}$) is the energy of VBM (CBM), respectively. $m^*_{VBM}$ ($m^*_{CBM}$) is the effective mass of VBM (CBM), respectively. For direct-gap semiconductor, e.g. MoS$_2$, $\mathbf{k}_{gap}$ is the $k$-vector of band gap position. In the expression of $H_{RKKY}$, the band energy can be written as

$$\varepsilon(\mathbf{k}) \approx E_{VBM} - m^*_{VBM} k^2 \qquad \text{for } \varepsilon(\mathbf{k}) < E_F$$

$$E_{CBM} + m^*_{CBM} k^2 \qquad \text{for } \varepsilon(\mathbf{k}) > E_F.$$

Then, $H_{RKKY}$ reads

$$H_{RKKY}(\mathbf{R}_i - \mathbf{R}_j) = -\mathbf{S}_i \cdot \mathbf{S}_j \frac{J^2 A_0^2}{(2\pi)^4} \int_{\substack{\varepsilon(\mathbf{k}) < E_F \\ \varepsilon(\mathbf{k}') > E_F}} \frac{e^{i\mathbf{k}'|\mathbf{R}_i-\mathbf{R}_j|\cos\theta'} e^{-i\mathbf{k}|\mathbf{R}_i-\mathbf{R}_j|\cos\theta}}{\varepsilon(\mathbf{k}') - \varepsilon(\mathbf{k})} k \, dk \, d\theta \, k' \, dk' \, d\theta'.$$

Next, a math tool would be imported. Starting from the expression $e^{ix\cos\theta} = 1 + ix\cos\theta - x^2\cos^2\theta/2! - ix^3\cos^3\theta/3! + x^4\cos^4\theta/4! + \ldots$, we have

$$\int_0^{2\pi} e^{ix\cos\theta} d\theta = \sum_{n=0}^{\infty} \frac{(-1)^n x^{2n}}{(2n)!} \int_0^{2\pi} \cos^{2n}\theta \, d\theta$$

$$= 2\pi \sum_{n=0}^{\infty} \frac{(-1)^n x^{2n}}{((2n)!!)^2}$$

$$= 2\pi \sum_{n=0}^{\infty} \frac{(-1)^n}{(n!)^2} \left(\frac{x}{2}\right)^{2n}$$

$$= 2\pi J_0(x).$$

Here $J_0(x)$ is zero-order Bessel function. Finally, the integral in $H_{RKKY}$ can be simplified as

$$H_{RKK}(\mathbf{R}_i - \mathbf{R}_j) = -\mathbf{S}_i \cdot \mathbf{S}_j \frac{J^2 A_0^2}{(2\pi)^4} \int_{\substack{\varepsilon(\mathbf{k}) < E_F \\ \varepsilon(\mathbf{k}') > E_F}} \frac{kk' J_0(k|\mathbf{R}_i - \mathbf{R}_j|) J_0(k'|\mathbf{R}_i - \mathbf{R}_j|)}{\varepsilon(\mathbf{k}') - \varepsilon(\mathbf{k})} dk \, dk'.$$